RESEARCH ARTICLE

# Personalized Recommender System for Children's Book Recommendation with A Realtime Interactive Robot

**Yun Liu, Tianmeng Gao, Baolin Song and Chengwei Huang***
*Fandou Information Technology Co. Ltd., Nanjing,210000, Jiangsu,China,{liuy,gaotm,songbl,huangcw}@fandoutech.com.cn.*

Corresponding Author: *Fandou Information Technology Co. Ltd., Nanjing,210000, Jiangsu,China,huangcwx@126.com.*

**Abstract:** In this paper we study the personalized book recommender system in a child-robot interactive environment. Firstly, we propose a novel text search algorithm using an inverse filtering mechanism that improves the efficiency. Secondly, we propose a user interest prediction method based on the Bayesian network and a novel feedback mechanism. According to children's fuzzy language input, the proposed method gives the predicted interests. Thirdly, the domain specific synonym association is proposed based on word vectorization, in order to improve the understanding of user intention. Experimental results show that the proposed recommender system has an improved performance and it can operate on embedded consumer devices with limited computational resources.

**Keywords:**personalized search; word vectorization; recommender system;children's robot

## 1. Introduction

Recommender systems are widely used to recommend products and services to end consumers.Various studies have been reported on building a recommender system. It is a growing field attracting more and more researchers from various backgrounds.

Recommender system is very important in consumer electronics. (Cheung et al.,2009), used comments from an online Chinese discussion forum to evaluate the credibility of recommendations, as an extension for conventional word-of-mouth study. In their work, how readers would be influenced by online recommendations was studied. The quality of recommendation played an important role.(Chen et al.,2004) from Carnegie Mellon University also presented similar influence of online recommendations on consumer sales. Their work showed that recommender systems could improve sales at Amazon.com.User comments can be collected and used for recommendation by text mining techniques(Aciar et al.,2007;Benlian et al.,2012). Kumar and Benbasat(Kumar & Benbasat,2006) studied business-to-consumer website, and recommender system had improved both perceived usefulness and social presence.(Lee et al.,2010), studied program recommendations in Digital Television. Refining the TV program selecting and satisfying the consumer's viewing requirements had a lot similarity with Children's book recommendation. Both requires adaptation and realtime response. In their work, many factors were considered to make good recommendation, such as: viewing pattern for contents, statistical information of viewing patterns, user's profile, etc.

A novel approach to recommend music to consumers, have been proposed by (Ayata et al.,2018). User emotions were detected based on physiological signals and feeded to a recommendation engine for improved consumer satisfaction.

These works have already revealed the usefulness of recommender system in many aspects of consumer electronics, but the application in educational robots have not been fully studied, especially for the recommendation of children's books in an child-robot interactive environment. In our work, specific factors related to children's reading interests are analyzed and used for recommendation.

On the other hand, interactive AI has been drawing a lot of research attentions and it leads to many applications in intelligent toys.(Park et al.,2019) investigated interactive AI for education purposes. In their work, virtual environment was created for language learning. Various efforts have been made for children's education and companion through interactive AI.(Ahn et al.,2018 )introduced an interactive toy for children to learn language. Their system used image recongnition technology to response with the name of an object. It was also running on embedded system, but it was not connected with adequate online resources to facilitate the learning.heidt and Pulver from Germany proposed a novel cube-like intelligent toy(Scheidt & Pulver, 2019). It could be used for STEM teaching purposes. Sensors were available with trainable functionality for image classification, which was implemented on edge devices. However the interaction mainly took place between machines, without much high level communication with children. (Faria et al.,2020), proposed to use interactive games to assist the monitoring of children's brain activities. The interactive environment helped to elicit specific cognitive process.

Natural language processing (NLP) plays an important role in interactive AI.(Patel et al.,2019) presented a chatbot for college students. Through natual language interaction the robot could provide useful information to anyone who accessed the university website.(Stefnisson & Thue.2018), proposed a novel application in story telling. Using NLP modules the robot could help authoring the story contents. However the AI agent could only understand simple relations and synonym association was not considered to improve the language understanding.(Ping et al.,2015)studied conversational AI during live conversation interactions. Cognitive principles were considered, which brought more adaptive and personalized language understanding ability. However deeper meaning of the text was not explored, such as user intention and user interests.

In our work, recommender system is designed in an interactive environment, in which children can talk freely with an intelligent robot and be guided to suitable reading and listening materials. Our robot is connected with a large amount of online books and resources, so that it may provide high quality content for children's education. Furthermore, the user intention is considered for better recommendation.

The recommender system can be viewed as a search ranking system, where the input query is a set of user and contextual information, and the output is a ranked list of items (Cheng et al.,2016).Memorization and generalization are two different techniques within a recommender system. Memorization learns the frequently co-occurred factors in historical data and infers the same response at the same input. Generalization uses machine learning algorithms to learn the patterns in the historical data so that it can infer new responses at similar inputs. For instance, (Cheng et al.,2016)proposed a framework called wide and deep learning combined a linear model component and a neural network component. It has good results on Goole play for recommending apps to users using massive-scale commercial app data.Another popular algorithm in recommender systems is collaborative filtering (Schafer et al.,2007). It can be coupled with deep learning for better content information ranking (Wang et al.,2015)

**Figure 1**

**The Children's Interactive Robot**

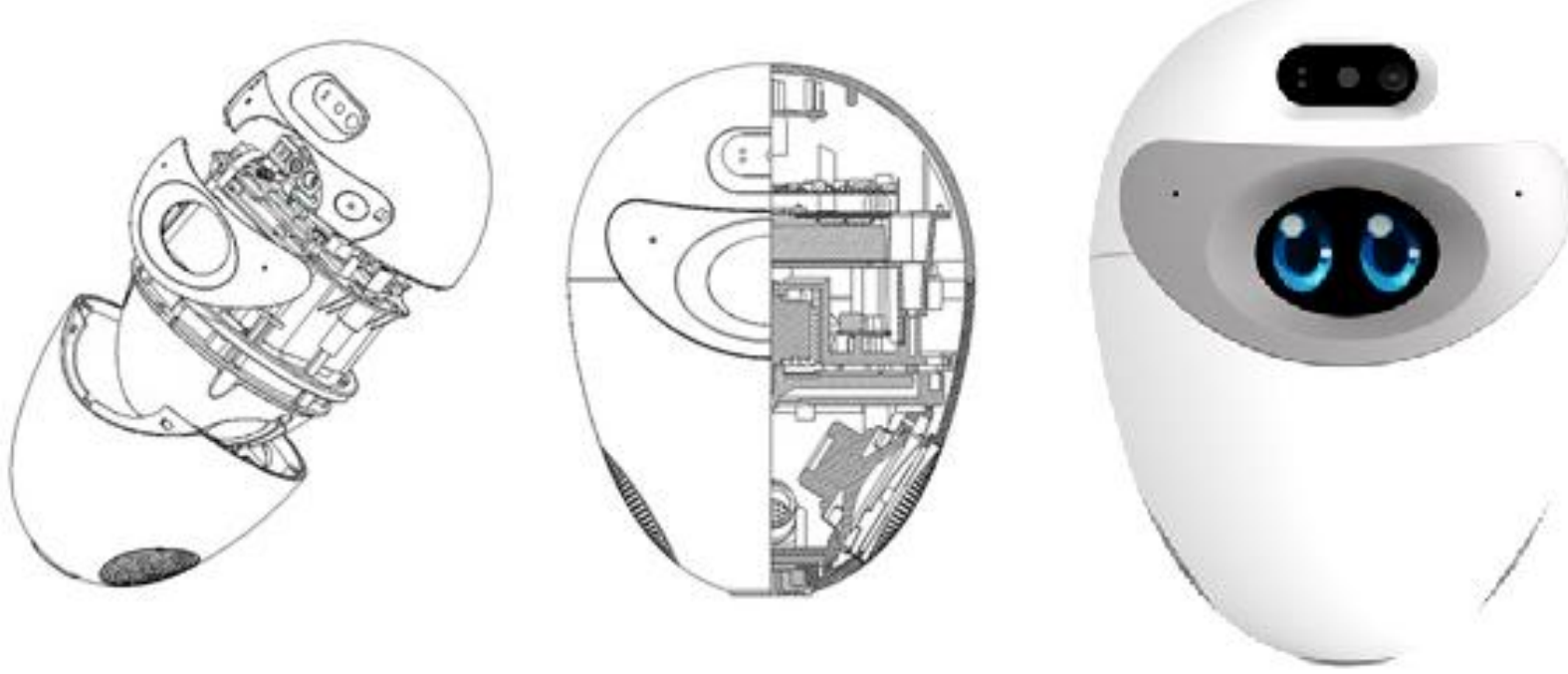

Although there are a few personalized search engines reported in the literature, the understanding of user intention remains a challenging problem. The following drawbacks in the conventional text search engine make it difficult to meet the actual needs of users.

First, computational cost of large scale text search is still high. With the rapid increase of online database, the computational burden on embedded systems is too high. When a stable wireless connection is not available, the computation costs of search algorithms are dependent on a local mobile device. In order to improve the quality and the speed of search engine, the efficiency of the conventional search algorithms need to be further improved.

Second, user interests are not fully considered in the search process. The keyword based text matching algorithms may not fully understand the user intention. Individual user's characters, personalities, search context, and interests should be considered in the search process. Searching is a passive way of recommendation, the matched text results should meet the actual needs of each individual user.

Third, semantic meaning of the large text data is not well represented in search algorithms. The text matching methods based on keywords are often lack of semantic understanding. It is difficult for the users to present every possible expression with the same meaning that they try to search in the text database. The recent developments in deep learning have resulted in an efficient representation of semantic meaning in words and phrases using word vectorization technique. When the search input is ambiguous and misleading, the search process becomes more difficult. Word vectorization may help to associate different keywords for matching. On the other hand, using natural language descriptions are more convenient than separate keywords and provides better user experience.

A word embedding method based on the deep learning is proposed by (Mikolov et al.2013;Mikolov et al.2013) named word2vec. When apply the neural network method to the conventional natural language model, the words can be represented as vectors in high dimensional space. The word embedding model is learned on a large text database. The relation between two words can be represented as the geometric distance between vectors. This provides a new perspective for text search systems. The difficulties in semantic understanding can be efficiently addressed by vectorization process. The keywords in search process can be associated to a list of related keywords that broaden the search rage and retrieve deeper information from text databases.

In this paper, we study the personalized recommender system with a realtime interactive robot for children, as shown in Fig.1. The robot can interact with users through speech and vision. A child user can tell the robot to start a reading session and what kind of books he or she is interested in. The robot needs to infer from the context and recommend books based on a child's words, which are usually ambiguous and inaccurate. The visual recognition ability enables the robot to understand the book content and tell a story(Huang & Jiang ,2019).

We propose a novel text search algorithm using a inverse filtering mechanism that is very efficient for labeled items search. The Bayesian network is used to implement the user interest prediction for an improved personalized search. It not only searches the related items using keywords, but also predicts user interests. The word vectorization is used during the process to discover potential targets with similar semantic meanings. The main contributions of our proposed system are three-fold: i)An efficient search algorithm based on Inverse Filtering(IF) is proposed; ii) A user interest prediction method based on Bayesian network is implemented, and combined with a personalized feedback mechanism; iii) Based on the word vectorization technique, a Domain Specific Synonym Association (DSSA) method is proposed to calculate the word distance in specific knowledge domain.

Key functionalities of our robot include, but not limited to, book reading , conversation, stories telling. Recommender system is one of the most important gateways for child users to learn with our robot. It separates this robot apart from other non-intelligent learning tools.

Since 2015, this robot, along with the proposed recommender system, has helped more than 10,000 homes in the consumer market and expected to continue growing. The algorithms proposed in this paper are implemented in the embedded systems of our robots, which are deployed in real-world environment satisfying realtime interaction demands.

The rest of the paper is organized as follows: Sec.2 provides related work on personalized recommender systems; Sec.3 gives the baseline methodology on fuzzy search and interest prediction, as well as the overall system architecture; Sec.4 introduces the details of DSSA method; experimental results are shown in Sec.5; finally, conclusions are drawn in Sec.6.

## 2. Related Work

During the realtime interaction between a robot and a user, key words can be used as an important clue for searching and recommendation. Full-text search engine is based on keyword matching technology, which can be used in realtime recommendation.Fuzzy search methods, KMP(Kunth et al.,1977) and BM(Boyer & Moore,1977) are proposed for approximate fast text matching.

However, a full-text search engine can't understand the user's deep semantics and needs, especially for fuzzy natural language input by the user.The traditional search engine returns the same result when different users enter the same query. Its search results are often inconsistent with the user's needs.

The personalized recommendation refers to the use of scenes, time and other information to understand user's real intention. The understanding of the natural language text is consistent with people's everyday thinking, therefore it can be used to build a smarter personalized recommender system(Guo et al.,2019).For instance, the natural language search engine Ask Jeeves (Kopytoff, 2010)supports for natural language search. Users could retrieve information through Q & A because there was a huge database of problems in the backend.

Several existing works studied book recommendation(Tewari et al.,2014; Rana & Jain,2012; Hariadi & Nurjanah,2017). (Tewari et al.,2014) investigated various approaches for book recommendation, including content filtering, collaborative filtering and rule based recommendation. Collaborative Filtering (Goldberg et al.,1992), is widely used in the existing online retail providers. In this algorithm, the users are given recommendations based on the similarity between its profile and other users' profiles. The similarity is defined on various metrics.(Rana & Jain,2012) looked into the application in online book shopping. In their work, a method with diverse recommendation was studied, and the recommended list might change over time.They adopted a temporal dimension that measures how much an item was liked by the user, which was time variant.( Hariadi & Nurjanah,2017)studied the personality in book recommendation. The personality factor in their work was built according to the similarity between users. The relation between user profile and user interest need to be further studied.

## 3. Personalized Recommender System for Children's Robot

We study the search algorithm in the context of children book recommendation. The books in the database are associated with text labels, such as names, authors, topics and other keywords. Matching these labels can be a straight forward way to search books. Mining the potential relations between these labels and a child's interest can provide better recommendation results.

### 3.1. Robot Architecture

The main chip has a four-core CPU whose frequency is up to 1.2GHz and a GPU that supports OpenGL. The main chip is connected with 1GB memory and 8GB flash drive space. The camera supports 5M pixel format and it is responsible for the visual interatction function. The audio codec supports 2-way speaker, 4-way microphone and 4-way earphone. They are responsible for the voice interaction function. The MCU is connected to the main chip with I2C, UART, and GPIO ports. It is responsible for motion control. The overall system power is 10W.

The software architecture of our robot is shown in Fig.2. The overall architecture consists of two layers, the platform layer and the application layer. In the application layer, we develope key functions including visual interaction, speech interaction, and language understanding. The visual interaction is introduced in our previous work(Huang &Jiang，2019). The speech interaction function involves realtime signal processing and enhancement. It is a natural and important mean to communicate with human(Huang et al., 2016). Different noise backgrounds are first detected and speech end point segmentation is carried out for preprocessing. Finally, the text content of the speech signal is generated using a local ASR engine or a cloud based ASR service, depending on the network connection quality.

Children's picture books are recognized and read to children. Before the robot can tell a story about any book, the recommender system is run and proposes books of interest. The recommender system is build together with other language modules, such as chatbot module, command understanding module. In the platform layer, JNI(Java Native Interface) is used to compile C++ libraries for faster computing.

**Figure 2**

**The SW Architecture of Our Interactive Robot**

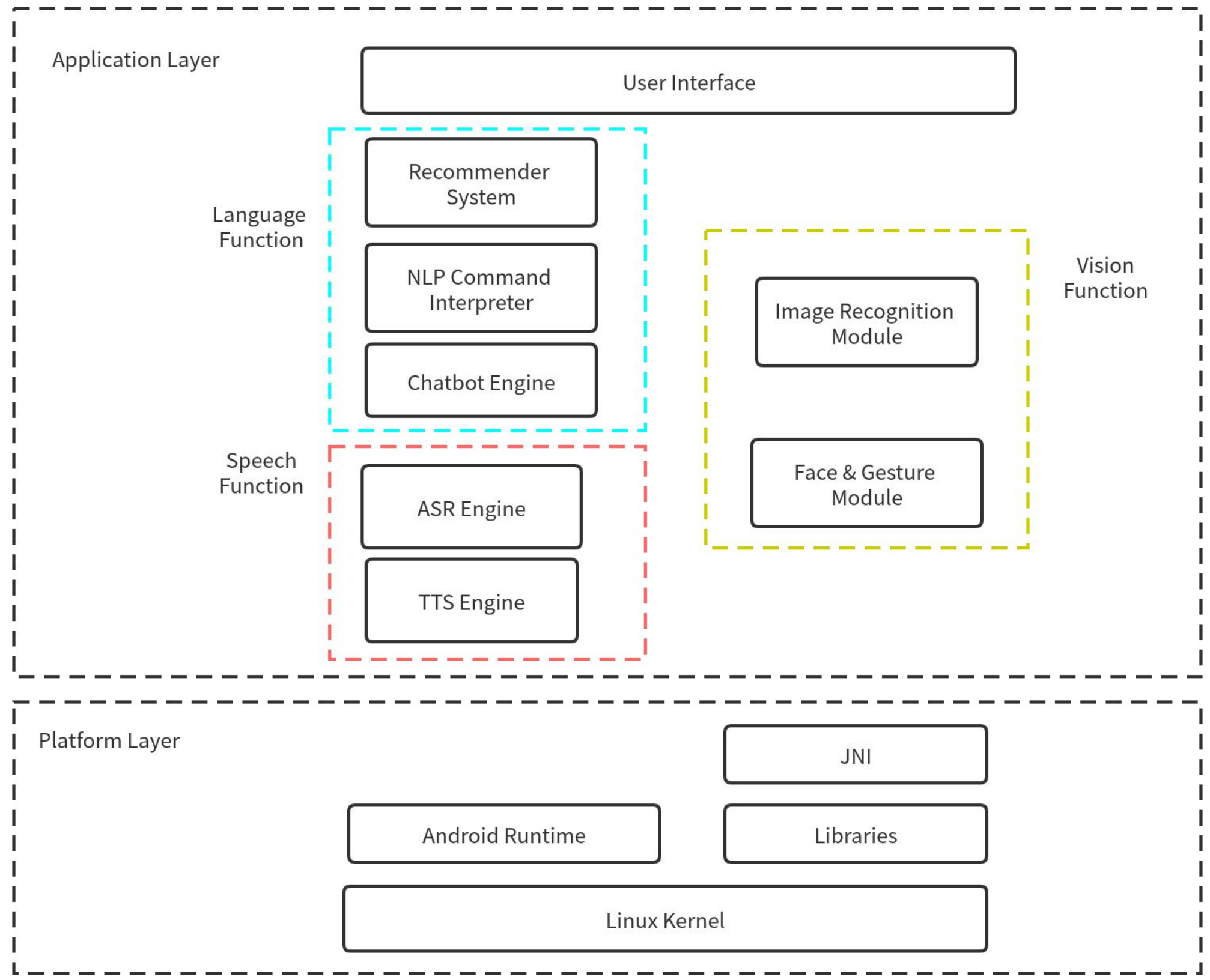

## 3.2. Approximate String Match Using Inverse Filtering

Various search algorithms can be used for text matching(Chen et al., 2009;Fu et al., 2015), but text matching can be time consuming. In this section, we propose to use a method called inverse filtering (IF) to accelerate the search process. The philosophy behind this method is based on two parts, the inverse part and the filtering part. The inverse idea refers to that, instead of going through a large amount of text to find a match with any keyword in the conventional approximate string match methods, we go though a small list of user input text contains some keywords to find matches with a large amount of book labels in the database. The filtering idea refers to that, we extract a very small alphabet from the user input text and use it to filter out any label that does not contains the characters in the alphabet. A predefined hash data structure makes this filter process very fast and the alphabet extraction works very well for Chinese language which is based on a large number of characters. Algorithm details will be given in the remaining of this section.

The inverse and filtering idea has been adopted previously in acoustic signal processing(Rothenberg,1973).

The IF based search algorithm mainly includes three steps as described below.

**Step One**: Building the hash index of resource database.

Before we search any children book, we label them using names, authors, topics, etc. Based on the book content we extract representative labels automatically using TF-IDF algorithm (Qaiser & Ali,2018; Salton & Buckley,1988]. In practice, we manually reviewed these labels on approximately 3,126 books for a better quality. TF-IDF algorithm is used to determine the weight of every keyword in the document and the high ranking words are considered as the representative labels. In this paper, the phrases book labels and book keywords refer to the same meaning, and label and keyword may be used interchangeably in the rest of the paper.

One fact about the Chinese language is that the number of characters in commonly used words and sentences are limited to a small number, approximately independent of the size of the book database. The number of characters tends to be stable when the number of keywords increases. For a Chinese database containing about $1.0\times10^4$ keywords, the number of characters is only around $3.5\times10^2$. So a hash map can be used to filter out the irrelevant keywords quickly.

Due to this property, the hash map is establishment before hand. A character may be contained by a keyword, and a keyword may be contained by a book. Therefore the first hash map $D_1$ is a map from any one character to a list of

keywords that contains it. The second hash map $D_2$ is a map from any one keyword to a list of books or items in the database that contains it.

**Step Two**: Match keywords using inverse filtering. The process of IF algorithm can be divided into two sub-steps.

First, go through the input text $X$ entered by the user in one search action and for each character $C$ in $X$ use the hash map $D_1$ to pick out the related keywords to build a candidate set $L_{cand}$ This process lies in the heart of IF, and it accelerates the search process significantly.

Second, go through all the keywords in $L_{cand}$ and for each keyword using a sliding window on the input text $X$ to find a approximately matched string if it exist. Dynamic time wrapping is used to compare sub-strings with different length and calculate the similarity scores. If the matching score is above threshold, the corresponding keyword is put in the hit label set $L_{hit.}$

**Step Three**:Post-process for ranking results.

When the input text provided by the user contains the exact expression of the targets in the database, the matched results are accurate and usually satisfactory. When the input text is ambiguous or misleading, a series of fuzzy results will be bit by the IF search algorithm. A post process for ranking the similarity score in the hit set $L_{hit}$ is necessary. The definition of the similarity score is shown in Eq.1, index i = 1, 2,..., N is the number of the keyword, and the distance between the keyword and a string contained in the sliding window is denoted as $dist_{DTW}(Key(i,m),X)$. The index m refers to different books or items in the database.

**Figure 3**
**The baseline framework of personalized fuzzy search system**

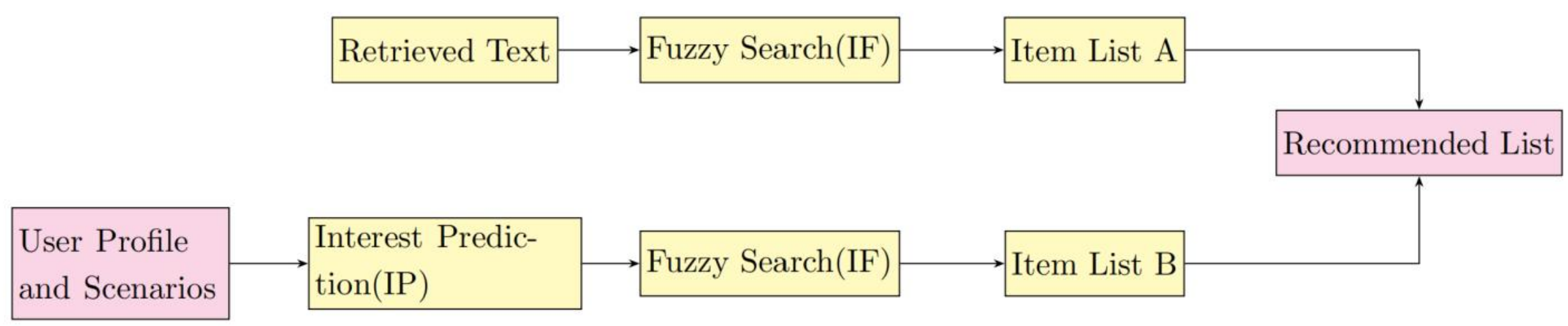

$$score(m) = \sum_{i=1}^{N} \frac{length(Key(i,m))}{dist_{DTW}(Key(i,m),X)} \tag{1}$$

Using hash map $D_2$ we can find the books related to any keyword. When a keyword is matched we can find out which book it belongs to. The results is ranked according to the scores and returned to the user.

## 3.3. IF Based Fuzzy Search Framework

In this section we introduce the basic configuration of our recommender system. It consists of user interest prediction module, IF search module and a feedback mechanism. The extension of the basic engine with a word association module is introduced in Sec.4. The overall system flowchart is show in Fig.3. The design of this engine is suitable for various general purposes, but it has a special focus on the recommender system in a virtual library where users share their opinions and comments on children's books.

The IF based keyword search module works independently at first taking in input sentences explicitly said by a user. The search result is denoted as $L_{exp}$. which consists a list of book names (items). The user interest prediction module takes in several factors from the user profile, such as age, gender, hobbies. It also takes some environmental factors as the input, such as time, date, interactive state, etc. Our search engine is deployed on an intelligent robot for dialogue based book recommendation. Therefore the interactive states may be important to predict user's search intention, such as playing a game, learning English, etc. The predicted user interests are then converted into a string of keywords and IF search algorithm is applied once more to generate another search result denoted as $L_{exp}$. The final search result is a combination of these two lists. A set of empirical rules is used to merge the two lists: a) if list $L_{exp}$ contains high score matches they should have higher ranks; b) the common item from list $L_{IF}$ and $L_{pre}$ should have higher ranks; c) the sizes of the lists should be limited. We take a direct implementation of these rules in simple IF-THEN form.

The user interest prediction(IP) can be very helpful. For instance, the input of "I am thirsty" from a user, might trigger a response with "football", "drinks", "sports bar", etc., in the interest list $L_{pre}$ and "sports bar", "banana milk", etc., in the $L_{exp}$ list. A combination of the interests and match results will place "sports bar" at the front of the search result list.

Consider some human-machine interaction application, the user might provide useful feed-backs on the search results. These feed-backs can be converted into simple accept (success) or reject (failure) in binary forms. A feedback adaptive mechanism is proposed to improve the interest prediction module for different individual user. The longer the prediction module is used, the better it is adapted and personalized for the user.

## 3.4. User Interest Prediction with Feedback Mechanism

In this section we implement the user interest prediction module as described above using Bayesian network (Pearl,1985;Atoui et al.,2019). The directed graphical model is used for inference user interest based on related factors like age, gender, personalities, hobbies, environmental factors, scenes, etc.

The network model is demonstrated in Fig.4, input layer nodes denote the observed variables (user profile factors) and they are independent of each other: $X = \{X_1, X_2, ..., X_n\}$. $X_i$ represents a keyword describing user's age range, gender type, personality type, hobby, time, day, scene, or any related environmental factor. The predicted interests are represented as language variables: $Y = \{Y_1, Y_2, ..., Y_n\}$. Each keyword is dependent on the combination of all the input variables. Through a training process the conditional probability distribution is learned.

Denote the conditional probability distributions in the network as: $\theta = \{P(Y_j|X_i)\}$. Maximum of A Posteriori (MAP) algorithm is used to estimate the parameters (Bassett & Deride,2019). The node in the output layer is dependent on the parent nodes in the above layer.

Given $Y_j = k$, the posteriori probability is calculated as:

$$P\left(X_i = h \middle| Y_j = k\right) = \frac{\sum_{t\in\phi} I\left(Y_j^{(t)} = k\right) \sum_{q\in\Omega} I\left(X_i^{(q)} = h\right)}{\sum_{t\in\phi} I\left(Y_j^{(t)} = k\right) \times n_t} \tag{2}$$

**Figure 4**
**The network structure of user interest prediction**

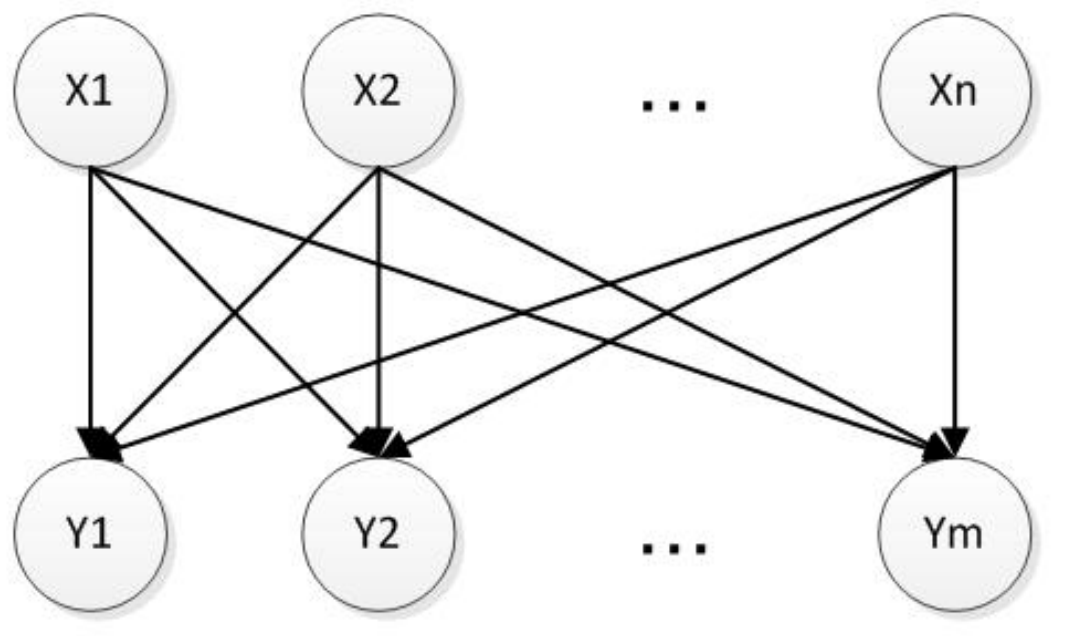

where $k$ denotes the interest index, $h$ denotes the factors in user profile, $t$ is an index of training samples, $\Phi$ is the set of training samples, $\Omega$ is the set of user profile factors, q is the index of the variables in the user profile set. I() is an index function, and if the expression is true it takes the value 1, otherwise it equals to 0.

In the above described Bayesian network based user interest prediction, an adaptive feedback adjustment mechanism is used to incrementally retrain the Bayesian network model. The longer a user uses this engine and gives feedback on the search results, the better it is personalized to this user. Feedback on success prediction will enhance the connection between the current inputs and outputs in the network, on the other hand, feedback on false search results will punish the prediction.

User's feedback is presented in binary form:

$$S = \begin{cases} 0 & reject \\ 1 & accept \end{cases} \tag{3}$$

The feedback mechanism of network parameters is based on the evaluation of the user's prediction results. The probability distribution θ=P($Y_j$|$X_i$) is updated in the current network model. $d_{new}$ is the incremental training data corresponded to the user feedback. It consists of the observable node set c$X_{observed}$, the predicted interest node $Y_j$ and the weight set $W$ between these probabilistic nodes. We have:

$$d_{new} = \Big\{ \big( X_1, X_2, \cdots, X_l, \mathrm{Y}_j, W_{1j}, W_{2j}, \cdots, W_{lj} \big) \Big| \\ \big( X_1, X_2, \cdots, X_l, \big) \in X_{observed}, \\ Y_j \in Y_R, \big( W_{1j}, W_{2j}, \cdots, W_{lj} \big) \in W \Big\}. \tag{4}$$

where $Y_R$ is the set of output variable. We update the weights as shown in Algorithm 1.

Algorithm 1 The incremental learning process of proposed feedback mech-anism.

Input: $W_{old}$

Output: $W_{new}$

1: Initialize: $W_{new} \leftarrow W_{old}$

2: for $X_i \in X_{observed} do$

3: weight $\xleftarrow{W_{ij} \in W_{old}} W_{ij}$

4: if S==1

5: weight=weight + th1

6: else

7: weight=weight - th2

8: end if

9: $W_{ij-new} \xleftarrow{W_{ij-new} \in W_{new}} weight$

10: end for

## 4. Domain Specific Synonym Association for Improved Recommendation

In order to understand the text semantic meaning in user input, word vectorization is used as a plug-in module to improve the above mentioned interest prediction and IF search. Specifically a domain specific synonym association method is used for user preference setting.

### 4.1. Domain Specific Synonym Association

Interest prediction may extend the search range on children book labels. However, some of the interest keywords are not directly linked to the labels if we don't understand the semantic meanings. Therefore, in this section we use the word vectors to associate one keywords to several of its nearest neighbors in the vector space. This process helps to understand the user's intention and expansions the search results.

A direct application of word2vec tool is able to associate some words to extend the search range, but we found that the neighboring words in the vectors space are not always contained in children books. Therefore, a domain specific word association method is needed to locate words within special topics.

The conventional word2vec model is retrained on a database of children books to make sure that the labels are represented in this vector space. Suppose there are $N$ words in total, and they are represented as vectors: $W=\{w_1,w2,\ldots,w_N\}$.

The book labels are also represented as K vectors: C={c1,c2,...,cK}. A spatial partitioning can be drawn with minimum mean distortion to classify wi to cj. Hence we can map the unrestricted words to a predefined set of book labels C.

Denote the dimension of a word vector is m. A word can be represented as wn = ( wn1,wn2,...., wnm ), where 1 ≤ n ≤N. A target word (book label) can be denoted as ck=(ck1,ck2,...,ckm), where 1 ≤k≤K.

For each vector $c_k$ in the target word set C, there is only one spatial region $S_k$ corresponded to it. The target word vector and the division of the K space is one to one mapping. Denote spatial regions as P={$S_1$,S2,..., $S_k$}, where $S_k$ corresponds to vector $c_k$. When a word vector $w_n$ is within the region of $S_k$, the mapping from the word vector $w_n$ to word vector $c_k$ can be obtained from Eq.5.

The mean square error can be used to measure the distortion, and it is defined as:

$$f: W \rightarrow C$$
$$f\left(w_n\right) = c_k, w_n \in S_k \tag{5}$$

The mean square error can be used to measure the distortion, and it is defined as:

$$\sigma_{ave} = \frac{1}{N} \sum_{n=1}^{N} \left\| w_n - f\left(w_n\right) \right\|^2 \tag{6}$$

In order to find the optimal set P={$S_1$,$S_2$,..., $S_K$}that minimizes the average error $\sigma_{ave}$, P needs to satisfy the nearest neighbor condition, i.e. the region $S_k$ should contain all the words closest to the vector $c_k$:

$$S_k = \left\{ w \middle| \left\| w - c_k \right\|^2 \leq \left\| w - c_{k'} \right\|^2, \forall k' = 1, 2, \cdots, K \right\} \tag{7}$$

By mapping the generally associated words to a domain specific set, reduces the keywords space by 10 times. It further improves the computational efficiency. The resulting synonym associated falls into the book labels set and the search using that synonym is more likely to hit a target book.

**Figure 5**
**Inproved IF module with domain specific synonym association**

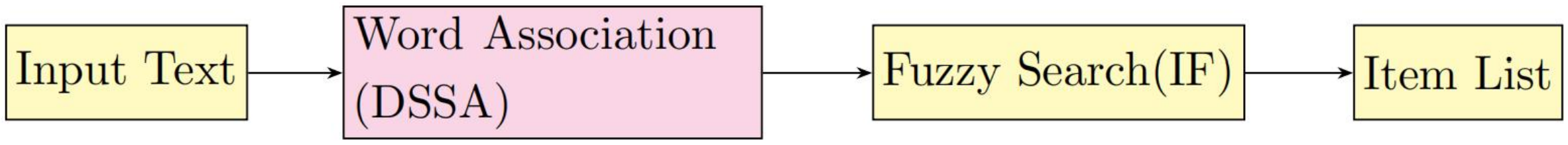

## 4.2. Overall Architecture

Using the domain specific synonym association(DSSA), the extended recommender engine with word2vec plug-in is depicted in Fig.5

Both the user input words and predicted interest keywords are extended by word2vec module, before send to IF based search module. The predicted interest descriptions are converted into keywords from the book labels in the database for an improved search experience, which will be demonstrated in Sec.5.

Using a retrained word2vec module as introduced in Sec.4.1, the keywords of the IF search module will be extended according to the semantic meaning of the input text. The associated keywords that generated by the word2vec module are based on the domain specific vocabulary of stories from children books. This plug-in module makes the search system more intelligent and more adaptive to children's needs.

## 5. Experimental Results

In our experiments, we verify the performance of our proposed recommender system in following four aspects. First, Sec.5.2 shows the results using only IF module. Second, Sec.5.1 shows the performance of our proposed novel IP module. Third, Sec.5.3 presents the baseline recommender system using IF module and IP module. Finally, in Sec.5.4, using the proposed novel DSSA module, the performance is compared with the conventional ones.

The following experimental results are tested on a PC with simulated environments for the convenience of reproducibility, except the embedded testing results in Tab.1. Nevertheless, all the data sets used are recorded from real-time interaction with children. All the algorithms are implemented on our embedded robot system.

## 5.1IF Based Fuzzy Search: Accuracy and Efficiency

The experiments on fuzzy search methods, KMP and BM are adopted for comparison with the proposed IF-based method. The algorithms are implemented using Java. The experiments on PC are carried out with 3.20 GHz CPU and 8GB memory. The tests on Robot are carried out with 1.2 GHz CPU and 1GB memory. The natural language processing toolkit FudanNLP(Qiu,2013) is used to assist the extraction of keywords from each book's content.

**Table 1**

**Comparison of three fuzzy search algorithms**

| Fuzzy Search Algorithms | Accuracy | Response Time (PC) | Response Time (Robot) |
|---|---|---|---|
| KMP | 89.29% | 4186ms | 1038ms |
| BM | 89.29% | 1679ms | 3977ms |
| IF | 91.27% | 607ms | 1521ms |

The search results are presented in book names. A search is considered successful when the result book names contain the predefined book (ground truth) that the user actually wants to find with a certain inquiry text (usually a sentence, or several sentences) he or she provided. The number of the book names returned by the search engine affects the success rate. The more books are allowed to present the higher the success rate goes, and the less smart the engine looks. In extreme condition, if we return all books in the database as the search result, the rate reaches 100%, suppose the targets are all included in the database. The inverse filtering based search, KMP matching algorithm and BM matching algorithm are compared in keyword search. The experimental results, tested both on PC and on our robot, are shown in Tab.1. The success of our keyword search test is defined differently from the above described book search. When a user provides an inquiry sentence that contains a keyword in an inexact form (missing a character or a wrong character), the search algorithms return a number of keywords matched. If the inquiry keyword is found among the returned keywords it is considered as a success.

We can see from Tab.1 that KMP based search and BM based search are the same in accuracy rates. The accuracy of the the proposed inverse filtering algorithm is higher than the other two conventional algorithms. The response time of KMP is the longest, and the IF method is significantly faster according to the tested response time. The efficiency of our search engine is improved by inverse filtering due to its hash map removing irrelevant candidate keywords.

**Figure 6**

**The precision rate of user interest prediction**

## 5.2User Interest Prediction Results

In order to verify the effectiveness of the user interest prediction method based on Bayesian network, we designed a test set that is consisted of two parts, the user profile and the user interest. The prediction problem is to find a map from user profile variables to user interest variables. Subjects are required to select 5 to 8 keywords that fit his or her profile and select 10 keywords from the a label set describing the common children's interests, e.g. football, video game, apple juice, reading, etc. This type of predicted interest key words may be considered as individual's intention during vocal interaction with a robot. We may infer the proper response based on the predicted key word. We use precision rate and recall rate (Kohavi,1995) to evaluate the performance of the prediction algorithm. K-fold cross validation is used to prevent over-fitting.

As shown in Fig.6 and Fig.7, the prediction results are represented as a number of interest keywords, and as the number of predicted keywords increases, the precision rate and the recall rate varies. The precision decreases when the number of predicted keywords increases, because some irrelevant words are introduced to the results. The recall rate increases when the number of predicted keywords increases, because more interest keywords are returned and it is less likely to miss the targets.

The training size to test size ratio starts from 70%, as we increase the training samples from user feedback and retrain the model incrementally, the performance is improved as observed in both precision rate and recall rate. Therefore the proposed feedback mechanism is effective.

**Figure 7**

**The recall rate of user interest prediction**

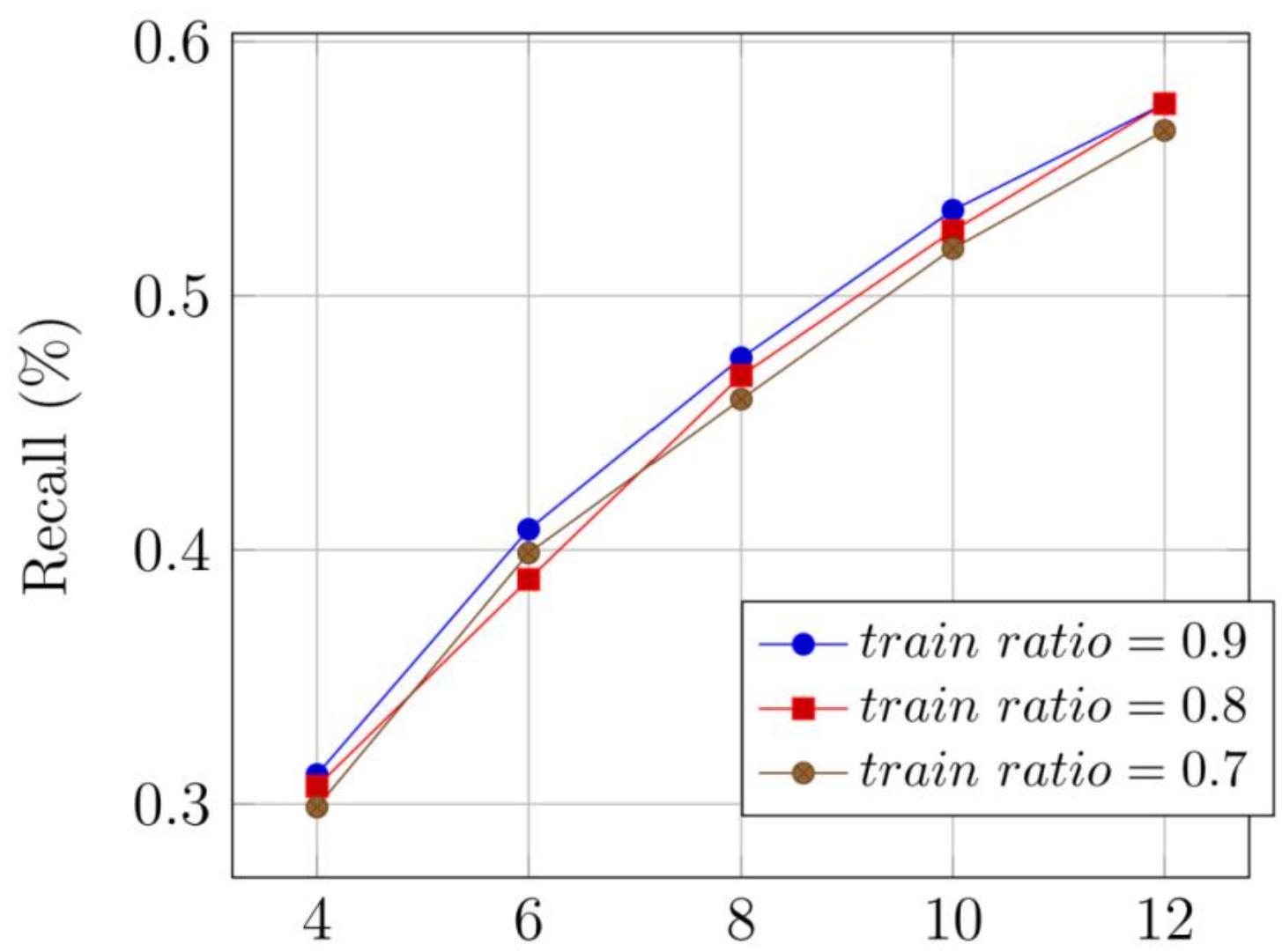

## 5.3Personalized Recommender System: Overall Performance

In this experiment we test the children book searching with different user profile setting. The user prediction and IF based search are integrated to perform the personalized fuzzy search. In the test we select 500 samples from 100 subjects. Those test samples contain the description of what the user wants to search for as the input text, the user profile described in separate keywords form, and the ground truth book name(s) the user actually wants to find as the target.

As show in Fig.8, the search accuracy is slightly improved in most of the settings, by using the user interest prediction method. When the number of returned book in the search result is set to 4 or 5, the improvements are more obvious.

Some examples in this search experiment are given in Tab.2. We can see that the search system can recommend books suitable for different users. This search results are therefore more personalized.

The reason behind the user interest prediction is that more keywords could be used in searching just knowing who the user is from the user profile without explicitly asking what the user what he or she wants to search.

**Figure 8**

**Search result with or without user interest prediction**

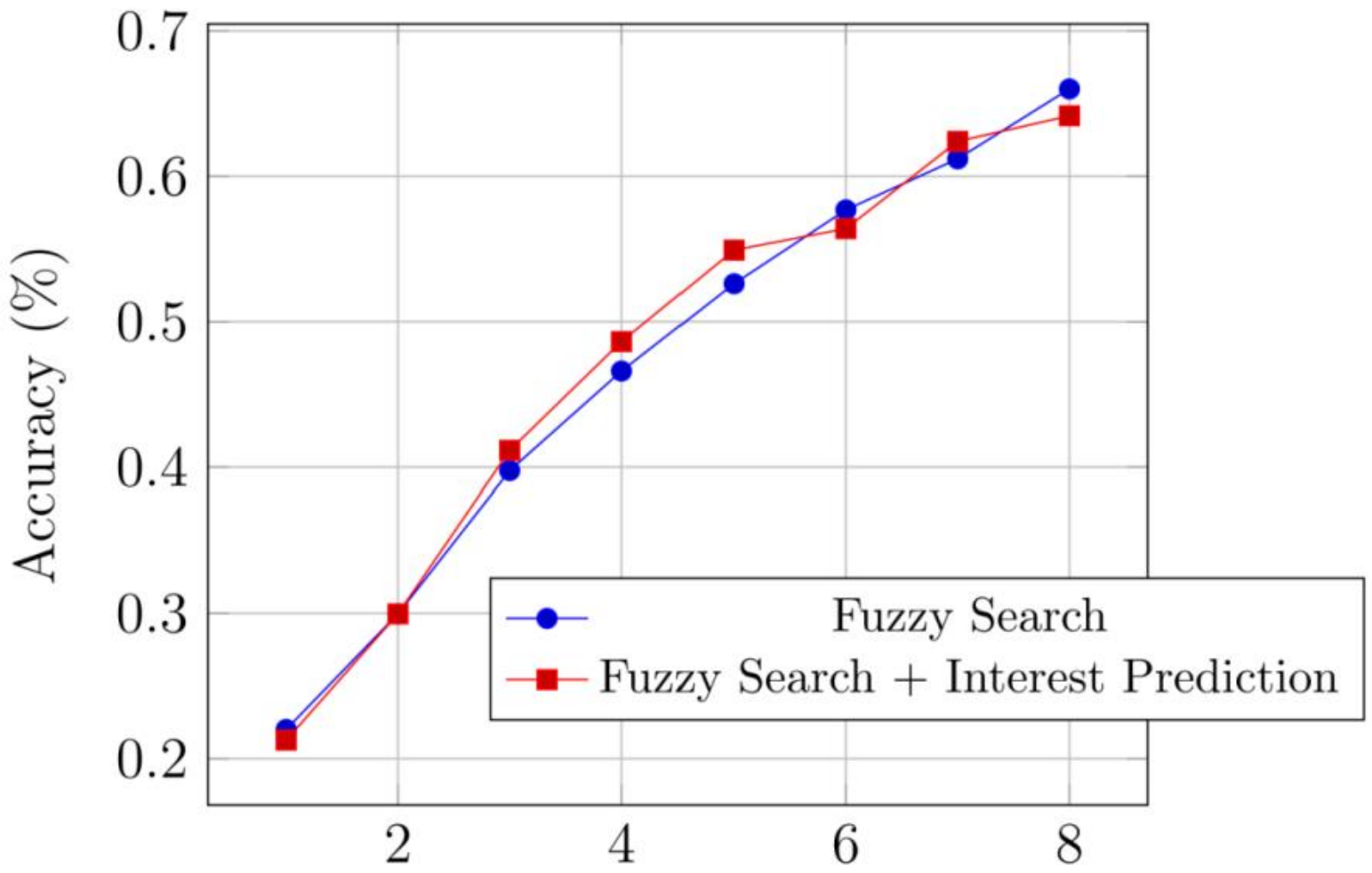

**Table 2**
**Examples of personalized search results(Translated in English)**

| Input Inquiry Text | User Profile | Predicted Interest Keywords | Search Result (book) |
|---|---|---|---|
| "Stories Related to Tom" | girl, night | sleep, moon, reading, piano | Tom's Nightmare |
| "Stories Related to Tom" | boy, summer | game, swimming, soccer, basketball | Tom Is in The Swimming Pool |
| "Stories Related to Emma" | boy, winter | game, soccer, snowman, basketball | Emma and Snowball Fight |
| "Stories Related to Emma" | girl, extroverted | reading, sleeping, communication, laughing | Emma and Her Friends |

**Table 3**
**The results of word vector association (Translated in English)**

| Words | Word2vec associative synonyms | Word2vec directional association synonym |
|---|---|---|
| music | play, singing, zhongshan, diction | play, sing, organ, poetry |
| mythology | Hebrew, orphans, Gang, Greek | Greece, Paris, history, astronomer |
| sun | other side, the past, 4 pieces, Moses | shine, go far, east, sky |
| butterfly | Primrose, intake, half, yellow | yellow, chrysanthemum, orchid, pink |

However, the problem of using this type of predicted interest labels is that the words and phrases we choose to describe a user's interests are not the same as what we choose to describe a book. Therefore, the search result in Fig.8 are not improved significantly. For instance, we may describe a user as ``fond of ping-pang'', when he may be looking for a book named ``fun games after school''.

A conventional way to bridge this gap is to use the same dictionary (a common set of words and phrases) when we label the books and setting user profiles. This may improve the search results directly. However, in this fashion, whenever we wants to apply the recommender engine to a different database or a different group of users, we need to change the dictionary and rebuilt every data hash. In this paper, as we proposed in Sec. 4, we plug-in word2vec module to overcome the difficulty in semantic understanding between ``fond of ping-pang'' and ``fun games after school'' in an efficient way.

### 5.4 Improved Personalized Recommender System: Domain Specific Synonym Association

In this section we test the performance using word association. The training corpus used for word2vec module is from the news corpus from Sogou laboratory(Wang et al.,2008) including 18 channels of news, e.g. domestic news, international news, sports news and entertainment news, recorded between June and July, 2015. When we apply the word association directly using this word2vec model, the results are shown in Tab.3.

It is obvious that the results are dependent on the content of the training corpus. The domain specific association can bring the results closer to book labels in the database. The final search performances are compared in Fig.9. Using synonym association we may provide more diversified response adapted to user intention, since the DSSA is based on the book content and user input.

**Figure 9**

**Compare the final Search results using improved DSSA module**

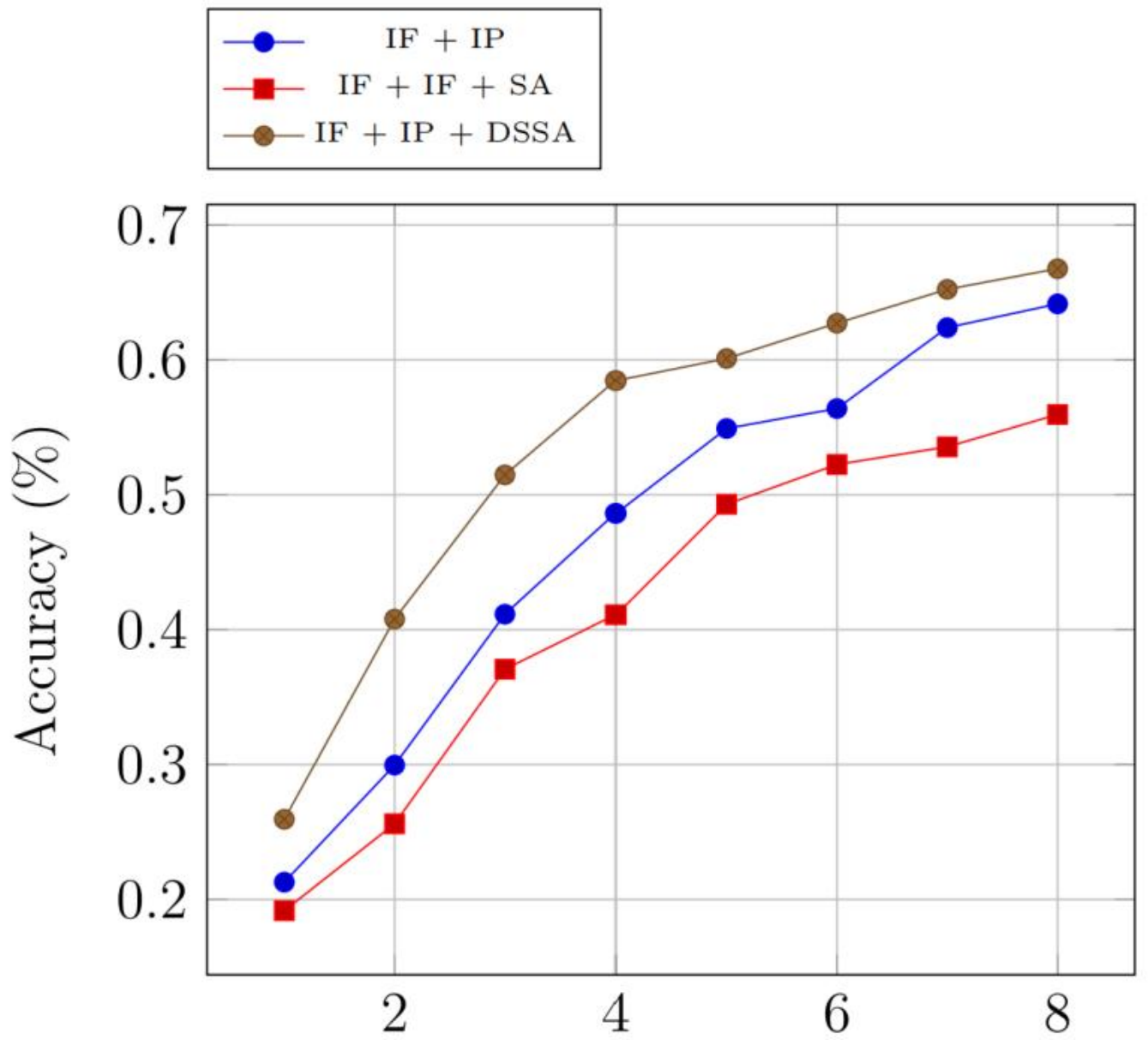

We can see from Fig.9 that the basic synonym association module does not improve the result. It associates the keywords to semantically close words and phrases in an open domain. The book labels are not more likely to be hit than any other words in a large corpus. By introducing more irrelevant keywords for searching decreases the result accuracy.

The use of domain specific synonym association, improved the final results significantly. The predicted words from user interest prediction and the associated synonyms are mapped onto the book labels in the semantic distance space created by word vectorization. Therefore the recommendation results are improved.

## 6. Conclusions

In this paper we study the personalized text search problem. The keyword based search method in conventional algorithms has a low efficiency in understanding users' intention since the semantic meaning, user profile, user interests are not always considered. We propose a novel text search algorithm using a inverse filtering mechanism that is very efficient for label based item search. Further more, we adopt the Bayesian network to implement the user interest prediction for an improved personalized search. According to user input, it searches the related items using keyword information, predicted user interest. In our approach, the word vectorization is used to discover potential targets according to the semantic meaning. Results show that the proposed recommender system has an improved efficiency and accuracy and it can operate on embedded consumer devices with very limited computational resources.

There are still several limitations in our proposed recommender system. Our search methods are only based on the short time interaction contains few rounds. Deeper understanding of multi-round interaction may improve the ability of intention prediction. The content we used for recommendation is limited to children's books. Extension to general knowledge and other learning materials using knowledge graph may be a valuable direction to study. Recommending the learning path and personalized learning plan is also out of our scope. In our future work, we will further investigate large language models to understand multi-round, multi-channel interaction where context information, facial expression, gestures can be considered for intention prediction.

## References

Aciar, S. , Zhang, D. , Simoff, S. & Debenham, J(2007). Informed recommender: Basing recommendations on consumer product reviews, *IEEE Intelligent Systems*, 22(3), 39-47.

Ahn, J. Y. , Kim, D. W. , Lee, Y. H. , Kim,W. , Hong, J. K. , Shim, Y. , et al.(2018). MOYA: Interactive AI toy for children to develop their language skills, *In Proc. of The 9th Augmented Human International Conference,* Seoul, Korea ,February, 2018,1-2.

Atoui, M. A. , Cohen, A. , Verron,S. ,& Kobi,A. (2019) .A single Bayesian network classifier for monitoring with unknown classes, *Engineering Applications of Artificial Intelligence*, (85), 681-690.

Ayata, D. , Yaslan ,Y. & Kamasak, M. E.(2018). Emotion based music recommendation system using wearable physiological sensors, *IEEE Transactions on Consumer Electronics,* 64(2), 196-203.

Bassett, R.,& Deride, J.(2019).Maximum a posteriori estimators as a limit of Bayes estimators, Mathematical Programming, 174(1-2), 129-144.

Benlian, A., Titah, R. & Hess, T.(2012). Differential effects of provider recommendations and consumer reviews in e-commerce transactions: An experimental study, *Journal of Management Information Systems*, 29(1), 237-272.

Boyer, R. S.,& Moore,J. S. (1977). A fast string searching algorithm, *Communications of The ACM*, 20(10), 762-772.

Cheung, M.Y., Luo, C., Sia, C. L., et al (2009). Credibility of electronic word-of-mouth: Informational and normative determinants of on-line consumer recommendations, *International Journal of Electronic Commerce*, 13(4), 9-38.

Chen, P. Y., Wu, S., & Yoon, J. (2004). The impact of online recommendations and consumer feedback on sales, *Proc. of International Conference on Information Systems*, 58.

Cheng, H. T. , Koc, L., Harmsen, J. ,et al.(2016). Wide & Deep Learning for Recommender Systems,*In Proc. of The 1st Workshop on Deep Learning for Recommender Systems*, Boston, MA, USA,September, 2016 ,7-10.

Chen, Y., Wang, W., Liu, Z., Lin, X.(2009) Keyword search on structured and semi-structured data. In Proceedings of the 2009 ACM SIGMOD International Conference on Management of data, 1005-1010.

Faria, D. R. , Bird, J. J. , Daquana, C. , Kobylarz ,J., & Ayrosa, P. P.(2020).Towards AI-based interactive game intervention to monitor concentration levels in children with attention deficit, *International Journal of Information and Education Technology*, 10(9), 641-648.

Fu, Z., Ren, K., Shu, J., Sun, X., Huang, F.(2015). Enabling personalized search over encrypted outsourced data with efficiency improvement. IEEE transactions on parallel and distributed systems, 27(9), 2546-2559.

Goldberg, D., Nichols, D. ,Oki B. M.,&Terry,D(1992). ``Using collaborative filtering to weave an information tapestry,'' Communication of The ACM, 35(12), 61-70.

Guo,W. , Gao, H. , Shi, J.,& Long, B.(2019) .Deep Natural Language Processing for Search Systems, *In Proc. of The 42nd International ACM Conference on Research and Development in Information Retrieval*, 1405-1406, Paris, France, July.

Hariadi,A. I., &Nurjanah,D. (2017).Hybrid attribute and personality based recommender system for book recommendation, *in Proc. of IEEE International Conference on Data and Software Engineering (ICoDSE)*, Palembang, South Sumatera, Indonesia,Nov., 2017, 1-5 .

Huang, C. , & Jiang, H.(2019). Image indexing and content analysis in children’s picture books using a large-scale database, Multimedia Tools and Applications, 78(15), 20679-20695.

Huang, C., Song, B., & Zhao, L. (2016). Emotional speech feature normalization and recognition based on speaker-sensitive feature clustering. International Journal of Speech Technology, 19, 805-816.

Knuth, D. E. , Morris, J. H., &Pratt, V. R.(1977).Fast pattern matching in strings, *SIAM Journal on Computing*, 6(2), 323-350.

Kopytoff, Verne G. (2010). Ask.com Giving Up Search to Return to Q-and-A Service .*The New York Times*.

Kohavi, R.(August 20-25, 1995). A study of cross-validation and bootstrap for accuracy estimation and model selection,*in Proc. of International Joint Conference on Artificial Intelligence*, 14(2), 1137-1145, Montreal, Quebec, Canada, .

Kumar, N. & Benbasat, I.(2006). Research note: the influence of recommendations and consumer reviews on evaluations of websites, Information Systems Research, 17(4) , 425-439.

Lee, S. , Lee, D & Lee, S. (2010). Personalized DTV program recommendation system under a cloud computing environment, *IEEE Transactions on Consumer Electronics*, 56(2), 1034-1042 .

Mikolov, T. , Chen, K., Corrado, G. , et al.(2013). Efficient estimation of word representations in vector space, *arXiv preprint, arXiv*:1301.3781.

Mikolov, T. , Sutskever, I. ,Chen,K. et al.(2013).Distributed representations of words and phrases and their compositionality, *Advances in neural information processing systems*, (26), 3111-3119.

Park, W. , Park, D. , Ahn, B., Kang, S. , Kim, H. , Kim, R. , & Na, J.(2019) . Interactive AI for linguistic education built on VR environment using user generated contents, *In Proc. of The 21st International Conference on Advanced Communication Technology (ICACT)*, Pyeongchang, Korea,February, 2019, 385-389.

Patel, N. P. ,Parikh, D. R., Patel,D. A. , &Patel, R. R. (June, 2019). AI and Web-Based Human-Like Interactive University Chatbot (UNIBOT), *In Proc. of The 3rd International Conference on Electronics, Communication and Aerospace Technology(ICECA)*, 148-150, Coimbatore, India .

Pearl,J. (1985). Bayesian networks: A model of self-activated memory for evidential reasoning, *in Proc. of The 7th Conference of the Cognitive Science Society*, 329-334, Irvine, CA.

Ping, Q., Niu, F. ,Thattai, G., Chengottusseriyil, J., Gao,Q. ,Reganti, A., et al.(2020). Interactive Teaching for Conversational AI, *arXiv preprint, arXiv*:2012.00958.

Qaiser ,S.,& Ali, R. (2018). Text mining: use of TF-IDF to examine the relevance of words to documents, International Journal of Computer Applications, 181(1), 25-29.

Qiu, X., Zhang, Q.,& Huang,X.(August, 2013). FudanNLP: A Toolkit for Chinese Natural Language Processing, i*n Proc. of The 51st Annual Meeting of the Association for Computational Linguistics: System Demonstrations*, Sofia, Bulgaria.

Rana,C. , &Jain, S. K. (2012). Building a Book Recommender system using time based content filtering, *WSEAS Transactions on Computers*, 11(2), 2224-2872.

Rothenberg, M. (1973). A new inverse‐filtering technique for deriving the glottal air flow waveform during voicing, *The Journal of The Acoustical Society of America*, 53(6), 1632-1645.

Salton, G. ,& Buckley, C.(1988). Term-weighting approaches in automatic text retrieval, *Information processing &management,* 24(5),513-523.

Scheidt, A., & Pulver, T. ( 2019). Any-cubes: a children's toy forlearning AI: Enhanced play with deep learning and MQTT, *In Proc. of Mensch und Computer*, Hamburg, Germany, September, 2019, 893-895.

Schafer, J. B., Frankowski, D. , Herlocker, J. , &Sen, S.(2007). Collaborative filtering recommender systems, The Adaptive Web, Heidelberg, Germany: Springer, 291-324.

Stefnisson,I. S. ,& Thue, D. (2018).Mimisbrunnur: AI-Assisted Authoring for Interactive Storytelling, *In Proc. of The Annual Artificial Intelligence and Interactive Digital Entertainment Conference(AIIDE)*, 236-242, AB, Canada, September.

Tewari, A. S. , Kumar, A. , &Barman,A. G.(2014). Book recommendation system based on combine features of content based filtering, collaborative filtering and association rule mining, *In Proc. of IEEE International Advance Computing Conference(IACC)*, Gurugram, Haryana, India, Feb. 2014,500-503.

Wang,C. , Zhang, M., Ma, S., et al.(2008), Automatic online news issue construction in web environment, *in Proc. of ACM International Conference on World Wide Web*, Beijing, China.

Wang, H. ,Wang,N. , & Yeung, D. Y.(2015). Collaborative deep learning for recommender systems, I*n Proc. of The 21th ACM International Conference on Knowledge Discovery and Data Mining*, 1235–1244, Sydney, NSW, Australia, August.

## Authors

**Yun Liu** received his Bachelor's Degree in Network Engineering from Nanjing University of Posts and Telecommunications in 2014, and Master's Degree in Information and Communication Engineering from School of Information Science and Engineering, Southeast University in 2017. He joined Fandou Technology as an algorithm engineer working on interactive robotics for children. His research interests include natural language understanding, machine learning and robotics.

**Tianmeng Gao** received her Bachelor's Degree in Information Management and Information System, from China Pharmaceutical University in 2012, and Master's Degree in Computer Science from School of Computer Science and Engineering, Southeast University in 2017. She joined Fandou Technology as an algorithm engineer focusing on interactive robotics for children. Her research interests include natural language understanding and deep learning.

**Baolin Song** recieved his B.S. and M.S. degrees from Qingdao University (QDU) and Shandong Normal University (SNU), in 2010 and 2014, respectively. He is currently working toward the Ph.D. degree at the School of Information Science and Engineering, Southeast University, China. He joined Fandou Technology as a research intern working on machine learning applications. His research interests include computer vision, pattern recognition, and facial expression recognition.

**Chengwei Huang** received his undergraduate degree in 2006, and Ph.D. for speech emotion recognition from Southeast University (China) in 2013. His main research interests include affective computing, signal processing and data mining. He conducted research on human-computer interaction as an associate professor in Soochow University from 2013 to 2014. He joined Fandou Technology in 2015, focusing on interactive robots for children, and he is currently with Zhejiang Laboratory focusing on intelligent computing. He can be contracted through huangcwx@126.com.